\documentclass[11pt]{article}
\usepackage{amssymb,amsmath}

\begin{document} 
\title{Equivalence between two-dimensional alternating/random Ising model  
and \\the ground state of \\one-dimensional alternating/random XY chain.} 
\author{Kazuhiko MINAMI}
\date{9/12/2012}
\maketitle
\abstract{ 
It is derived that the two-dimensional Ising model with alternating/random interactions and with periodic/free boundary conditions is equivalent to the ground state of the one-dimensional alternating/random XY model with the corresponding periodic/free boundary conditions. 
This provides an exact equivalence between a random rectangular Ising model, 
in which the Griffiths-McCoy phase appears, and a random XY chain. }
\\
\vspace{2.4cm}

\noindent
\vspace{0.3cm}

\noindent
Graduate School of Mathematics, Nagoya University,\\
Nagoya, 464-8602, JAPAN
\vspace{0.3cm}

\noindent
minami@math.nagoya-u.ac.jp

\newpage
Let us consider the two-dimensional rectangular Ising model 
\begin{eqnarray}
H_{\rm I}=-\sum_{i=1}^M\sum_{j=1}^N[
J_{j}^{h}\sigma_{ij}^x\sigma_{ij+1}^x+J_{j}^{v}\sigma_{ij}^x\sigma_{i+1j}^x],
\label{2DIsing-Hamiltonian}
\end{eqnarray}
where $\{\sigma_{ij}^k\} \:(k=x, y, z)$ are the Pauli operators 
satisfying $\sigma_{i+Mj}^k=\sigma_{ij}^k$ and $\sigma_{ij+N}^k=\sigma_{ij}^k$. 
The partition function of the model (\ref{2DIsing-Hamiltonian}) is obtained 
from the maximum eigenvalue of the transfer matrix 
\begin{eqnarray}
V=V_1^{1/2}V_2V_1^{1/2},
\label{transfermatrixV}
\end{eqnarray}
where
\begin{eqnarray}
V_1&=&\left[\Pi_{j=1}^N\frac{e^{K_j^{v}}}{\cosh K_{j}^{v*}}\right] \exp[\sum_{j=1}^NK_{j}^{v*}\sigma_{j}^z],
\nonumber\\
V_2&=&\exp[\sum_{j=1}^NK_{j}^{h}\sigma_{j}^x\sigma_{j+1}^x],\hspace{0.3cm}
K_{j}^{l}=J_{j}^{l}/k_BT \:\:(l=v, h).\: 
\nonumber
\end{eqnarray}
Here $\{\sigma_{j}^k\}$ are again the Pauli operators 
satisfying $\sigma_{j+M}^k=\sigma_{j}^k$,   
and $K_{j}^{v*}$ is the dual interaction of $K_{j}^{v}$ defined by 
$\tanh K_{j}^{v*}=\exp(-2K_{j}^{v})$. 

Suzuki derived\cite{71SuzukiLetter} \cite{71Suzuki} that 
the transfer matrix (\ref{transfermatrixV}) 
with the uniform interactions $K_{j}^{v}=K^{v}$ and $K_{j}^{h}=K^{h}$
commute with the Hamiltonian of the one-dimensional quantum XY model
\begin{eqnarray}
H_{\rm XY}=-\sum_{j=1}^N[
J_j^x\sigma_{j}^x\sigma_{j+1}^x+J_j^y\sigma_{j}^y\sigma_{j+1}^y]
-\sum_{j=1}^N \mu H_j\sigma_{j}^z,
\label{1DXY-Hamiltonian}
\end{eqnarray}
with the uniform interactions $J_j^k=J^k \:(k=x, y)$ 
and the uniform external fields $H_j=H$.  
He showed that two operators $V$ and $H_{\rm XY}$ commute 
when the coupling parameters satisfy the condition that
\begin{eqnarray}
J^y/J^x=\tanh^2K^{v*},
\label{cond-uniform-JJ}\\
\mu H/J^x=2\tanh K^{v*}\coth 2K^{h}. 
\label{cond-uniform-H}
\end{eqnarray}
Hence $V$ and $H_{\rm XY}$ can be diagonalized 
with a common basis set of eigenvectors. 
In particular, the eigenstate for the maximum eigenvalue of $V$ 
coincides with the ground state of $H_{\rm XY}$, 
and hence the thermodynamic properties, 
such as correlation functions and critical singularities 
of the two-dimensional Ising model 
are related to those obtained in the ground state of the one-dimensional quantum XY model. 

In this letter, 
it is derived that this equivalence can be extended 
to the models with general coupling parameters 
including alternating and random interactions and external fields 
with periodic or free boundary conditions. 
The condition 
\begin{eqnarray}
[V, H_{\rm XY}]=0,
\label{comrel}
\end{eqnarray}
is equivalent to  
\begin{eqnarray}
V_2V_1^{1/2}H_{\rm XY}V_1^{-1/2}V_2^{-1}=V_1^{-1/2}H_{\rm XY}V_1^{1/2}. 
\label{comrel2}
\end{eqnarray}
One obtains
\begin{eqnarray}
V_1^{1/2}H_{\rm XY}V_1^{-1/2}=\hspace{5.5cm}
\nonumber\\
-\sum_{j=1}^N[
J_j^x
(c_j^+\sigma_{j}^x\sigma_{j+1}^x-c_j^-\sigma_{j}^y\sigma_{j+1}^y
+s_j^+i\sigma_{j}^y\sigma_{j+1}^x+s_j^-i\sigma_{j}^x\sigma_{j+1}^y)
\nonumber\\
+J_j^y
(-c_j^-\sigma_{j}^x\sigma_{j+1}^x+c_j^+\sigma_{j}^y\sigma_{j+1}^y
-s_j^-i\sigma_{j}^y\sigma_{j+1}^x-s_j^+i\sigma_{j}^x\sigma_{j+1}^y)
]
\nonumber\\
-\sum_{j=1}^N \mu H_j\sigma_{j}^z,\hspace{1.2cm}
\end{eqnarray}
where 
\begin{eqnarray}
c_j^\pm
&=&\frac{1}{2}[\cosh(K_{j}^{v*}+K_{j+1}^{v*})\pm\cosh(K_{j}^{v*}-K_{j+1}^{v*})]
\nonumber\\
s_j^\pm
&=&\frac{1}{2}[\sinh(K_{j}^{v*}+K_{j+1}^{v*})\pm\sinh(K_{j}^{v*}-K_{j+1}^{v*})].
\nonumber
\end{eqnarray}
The coefficient of the term $\sigma_{j}^y\sigma_{j+1}^y$ vanishes 
if we assume 
\begin{eqnarray}
J_j^xc_j^-=J_j^yc_j^+. 
\label{cond-general-1}
\end{eqnarray}
Then it is straightforward to derive that (\ref{comrel2}) is satisfied 
if we assume 
\begin{eqnarray}
\mu H_j &=&
\mu H_j {\tilde c}_j^+-{\tilde J}_{j-1}^-[+-]_j-{\tilde J}_{j}^+[-+]_j
\nonumber\\
0&=&
\mu H_j {\tilde c}_j^--{\tilde J}_{j-1}^-[-+]_j-{\tilde J}_{j}^+[+-]_j
\nonumber\\
-{\tilde J}_{j-1}^-&=&
(-\mu H_j)[+-]_j+{\tilde J}_{j-1}^-{\tilde c}_j^++{\tilde J}_{j}^+{\tilde c}_j^-
\nonumber\\
-{\tilde J}_{j}^+&=&
(-\mu H_j)[-+]_j+{\tilde J}_{j-1}^-{\tilde c} _j^-+{\tilde J}_{j}^+{\tilde c}_j^+,
\label{cond-general-2}
\end{eqnarray}
where 
${\tilde J}_{j}^\pm=s_j^\pm J_j^x-s_j^\mp J_j^y$, and 
\begin{eqnarray}
{\tilde c}_j^\pm
&=&\frac{1}{2}[\cosh(2K_{j-1}^{h}+2K_{j}^{h})\pm\cosh(2K_{j-1}^{h}-2K_{j}^{h})]
\nonumber\\
{\tilde s}_j^\pm
&=&\frac{1}{2}[\frac{\sinh(2K_{j-1}^{h}+2K_{j}^{h})}{2K_{j-1}^{h}+2K_{j}^{h}}\pm\frac{\sinh(2K_{j-1}^{h}-2K_{j}^{h})}{2K_{j-1}^{h}-2K_{j}^{h}}],
\label{def-coeff}
\end{eqnarray}
and the symbols $[+-]_j$ and $[-+]_j$ denote 
\begin{eqnarray}
[\pm, \mp]_j=2K_{j-1}^{h}{\tilde s}_j^\pm+2K_{j}^{h}{\tilde s}_j^\mp.
\nonumber
\end{eqnarray}
From the equation (\ref{cond-general-1}), $J_j^x$ and $J_j^y$ can be written as  
\begin{eqnarray}
J_j^x &=&(\cosh K_{j}^{v*})(\cosh K_{j+1}^{v*}) \:\tau_j,\nonumber\\
J_j^y &=&(\sinh K_{j}^{v*})(\sinh K_{j+1}^{v*}) \:\tau_j . 
\label{cond-JJ}
\end{eqnarray}
Therefore, $V$ and $H_{\rm XY}$ commute 
if (\ref{cond-general-2}) and (\ref{cond-JJ}) are satisfied. 

There are solutions for the equations (\ref{cond-general-1}) and (\ref{cond-general-2}) 
in the following cases:

1-1) Uniform interactions: 
The interactions of the Ising model are uniform, 
$K_j^v=K^v$ and $K_j^h=K^h$. 
In this case, $c_j^\pm=c^\pm$, $s_j^\pm=s$, 
${\tilde c}_j^\pm={\tilde c}^\pm$, and ${\tilde s}_j^\pm={\tilde s}^\pm$,
where
\begin{eqnarray} 
c^\pm&=&\frac{1}{2}[\cosh 2K^{v*}\pm 1], \hspace{0.3cm}
s=\frac{1}{2}\sinh 2K^{v*},\nonumber\\
{\tilde c}^\pm&=&\frac{1}{2}[\cosh 4K^{h}\pm 1], \hspace{0.3cm}
{\tilde s}^\pm=\frac{1}{2}[\frac{\sinh 4K^{h}}{4K^{h}}\pm 1]. \nonumber
\end{eqnarray}  
Assuming the uniform interactions $J_j^k=J^k \:(k=x, y)$ 
and the uniform external fields $H_j=H$, 
the equations  in (\ref{cond-JJ}) result in (\ref{cond-uniform-JJ}), and 
four equations in (\ref{cond-general-2}) 
reduce to one identical condition 
\begin{eqnarray}
\mu H=2s(J^x-J^y)\frac{\sinh 4K^h}{\cosh 4K^h-1}, 
\label{cond-H}
\end{eqnarray}
which results in (\ref{cond-uniform-H}).  

1-2) Alternating-sign $K_j^h$: 
Let us consider the case 
in which 
the vertical interactions of the Ising model are uniform, $K_j^v=K^v$, 
and the horizontal interactions alternate in sign 
as $K_j^h=(-1)^j K^h$. 
In this case, $c_j^\pm=c^\pm$, $s_j^\pm=s$,  
${\tilde c}_j^\pm=\pm{\tilde c}^\pm$, and ${\tilde s}_j^\pm=\pm{\tilde s}^\pm$.  
Assuming $J_j^k=(-1)^j J^k \:(k=x, y)$, i.e. $\tau_j=(-1)^j \tau$, 
we again obtain (\ref{cond-H}). 

1-3) Alternating $K_j^v$: 
Let us consider the case 
in which the vertical interactions of the Ising model alternate as 
$K_j^v=K_{\rm A}^v$ for $j=$ odd and 
$K_j^v=K_{\rm B}^v$ for $j=$ even, 
and the horizontal interactions are uniform, $K_j^h=K^h$. 
In this case 
$c_j^\pm=c_{\rm AB}^\pm$, 
$s_j^\pm=s_{\rm AB}^\pm$ for $j=$ odd and 
$s_j^\pm=s_{\rm AB}^\mp$ for $j=$ even, 
${\tilde c}_j^\pm={\tilde c}^\pm$, and ${\tilde s}_j^\pm={\tilde s}^\pm$,  
where $c_{\rm AB}^\pm$ and $s_{\rm AB}^\pm$ are obtained from (\ref{def-coeff})  
substituting $K_j^v=K_{\rm A}^v$ and $K_{j+1}^v=K_{\rm B}^v$. 
Assuming  $\tau_j=\tau$, 
the equations in (\ref{cond-general-2}) reduce to one condition 
that results in alternating external fields 
\begin{eqnarray}
\mu H_j=2\:(\frac{1}{2}\sinh 2K_j^{v*})\: \tau\frac{\sinh 4K^h}{\cosh 4K^h-1}.  
\nonumber
\end{eqnarray}
In this case, $H_j$ are alternating for $j$ being even and odd, 
but $J_j^x$ and  $J_j^x$ are uniform, 
i.e. $J_j^x=J^x=(\cosh K_{\rm A}^{v*})(\cosh K_{\rm B}^{v*}) \:\tau$ 
and $J_j^y=J^y=(\sinh K_{\rm A}^{v*})(\sinh K_{\rm B}^{v*}) \:\tau$.  
We assume that $K_{\rm A}^v$ and $K_{\rm B}^v$ have the same sign 
so that the couplings of the XY model remain real   
(note that the dual interaction of $-K^v$ is $(-K^v)^*=K^{v*}\pm\pi i/2$).  

Two cases 1-2) and 1-3) are not exclusive. 
The Ising model with alternating-sign $K_j^h$ and alternating $K_j^v$ 
has its equivalent XY chain. 

1-4) Random-sign $K_j^h$: 
Let us consider the case 
where the horizontal interactions of the Ising model 
have the same absolute value but random signs,  
$K_j^h=\epsilon_j K^h$, where $\epsilon_j=\pm 1$. 
In this case we obtain 
$c_j^\pm=c^\pm$, $s_j^\pm=s$, 
and also obtain that  
${\tilde c}_j^\pm={\tilde c}^\pm$ and ${\tilde s}_j^\pm={\tilde s}^\pm$ 
for $\epsilon_{j-1}=\epsilon_j$, and 
${\tilde c}_j^\pm=\pm{\tilde c}^\pm$ and ${\tilde s}_j^\pm=\pm{\tilde s}^\pm$ 
for $\epsilon_{j-1}=-\epsilon_j$.  
Assuming the sign of the interactions as $J_j^k=\epsilon_j J^k \:(k=x, y)$, i.e. $\tau_j=\epsilon_j \tau$, 
we again obtain the solution (\ref{cond-H}). 

1-5) Arbitrary $K_j^v$: 
Let us consider the case 
where the vertical interactions of the Ising model $\{K_j^{v}\}$ 
have arbitrary strength with the same sign. 
A solution exists provided that ${\tilde J}_{j-1}^-={\tilde J}_{j}^+$ for each $j$.  
In this case, $\tau_j=\tau$ in (\ref{cond-JJ}) and the external fields are obtained as 
\begin{eqnarray}
\mu H_j=2s_j \tau\frac{\sinh 4K^h}{\cosh 4K^h-1},\hspace{0.2cm}
s_j=\frac{1}{2}\sinh 2K_j^{v*}. 
\label{1-5-H}
\end{eqnarray}

The cases 1-2) and 1-3) are special cases of 1-4) and 1-5), respectively. 
Again two conditions 1-4) and 1-5) are not exclusive. 
Assuming 1-4), $K_j^h=\epsilon_j K^h$ and $J_j^k=\epsilon_j J^k \:(k=x, y)$, 
a solution for 1-5) exists 
when ${\tilde J}_{j-1}^-= {\tilde J}_{j}^+$. 
The external fields are obtained from (\ref{1-5-H}). 

For all of these cases, from 1-1) to 1-5), a solution exists 
2-1) with the periodic boundary condition, and 
2-2) with the free boundary condition. 
Assuming that $K_N^h=0$,  
we obtain 
${\tilde c}_{N}^+=\cosh 2 K^h$,  ${\tilde c}_{N}^-=0$, 
${\tilde s}_{N}^+=\sinh 2 K^h/2 K^h$ and  ${\tilde s}_{N}^-=0$. 
Assuming a finite XY chain $\tau_N=0$, 
and hence ${\tilde J}_{N}^+=0$, 
the condition (\ref{cond-general-1}) for $j=N$ is satisfied with $0=0$, 
and (\ref{cond-general-2}) for $j=N$ have a solution 
\begin{eqnarray}
\mu H_N={\tilde J}_{N-1}^-\frac{\sinh 2K^h}{\cosh 2K^h-1}.  
\nonumber
\end{eqnarray}

Some of these systems 1-1)-1-5), with 2-1) or 2-2) 
have been introduced 
in connection with experiments and theoretical interests. 

Fisher and Ferdinand calculated\cite{67Fisher}
the free energies of two-dimensional Ising models 
on various lattices with various boundary conditions.  
Their motivation was to explain 
the shifts and rounding of the specific heat maximum observed in experiments, 
in terms of the effects of microcrystalline structure.  
They introduced a grain boundary in which 
a vertical ladder of $j$ horizontal bonds has modified interactions $\xi J$. 
Their case with $\xi=0$ and $j\to\infty$ 
corresponds to our case 1-1) with 2-2). 

The two-dimensional Ising model with the periodic boundary condition in the horizontal direction 
and the free boundary conditions in the vertical direction 
was considered in a study by Schultz et al.\cite{64SchultzMattisLieb},  
in which they diagonalized the model as a many-fermion system 
and obtained the free energy.  
Abraham solved\cite{71Abraham} 
the eigenvalue problem for the transfer matrix of the rectangular Ising model 
with the periodic boundary condition in the vertical direction  
and the free boundary conditions in the horizontal direction,  
and investigated the detailed band structure. 
This model corresponds to the case 1-1) with 2-2).

The Ising model on a cylinder 
with the boundary condition that 
the spins on one edge are fixed to be up and 
the spins on the other edge are fixed to be down\cite{71AbrahamGallavotti}, 
and that with another boundary condition that 
the spins on the two edges are fixed to be up\cite{73Abraham},  
were considered
in order to shed light on the surface tension problem 
between two oppositely magnetized phases.  

McCoy and Wu\cite{68McCoy} considered
the rectangular Ising model
with uniform horizontal interactions, 
and vertical interactions that vary randomly from row to row,
with the periodic boundary condition in the horizontal direction,
and with the free boundary conditions in the vertical direction.
They attempted to explain experiments that showed that
the specific heat is a smooth function at $T_c$.
They suspected that this effect was due to the presence of random impurities.
This model coincides with our case 1-4) with 2-2) 
if the distribution of the random interactions is assumed to be random $\pm J$,
though they assumed a narrow power law distribution function
in their calculations.
Their distribution function can be introduced in our case 1-5) with 2-2),
and it is not unreasonable to expect similar behavior in our system.
They obtained the result that
the specific heat is not divergent and infinitely differentiable at and near $T_c$
though it is not analytic.
It was also derived in another paper \cite{69McCoy} that
the usual parametrization in terms of critical exponents
does not describe the critical behavior of this model.

Shanker and Murthy\cite{87ShankarMurthyRC}\cite{87ShankarMurthy} considered a model
in which the vertical interactions are fixed and ferromagnetic,
and the horizontal interactions are equal in each row and vary randomly in sign and in magnitude.
Their model generally includes frustration.
They assumed the periodic boundary conditions in two directions.
This model corresponds to our case 1-5) with 2-1),
provided that the horizontal interactions are random but have the same sign.
They mapped the problem to a collection of one-dimensional random field problems,
and identified three phase transitions.

The commutation relation (\ref{comrel}) certifies 
that $V$ and $H_{\rm XY}$ can be diagonalized simultaneously. 
Moreover, it can be derived that 
the eigenstate for the maximum eigenvalue of $V$ coincides with 
the ground state of $H_{\rm XY}$, 
and hence the thermodynamic properties of the two-dimensional Ising models 
and the ground state properties of the one-dimensional XY models 
are related to each other. 

In order to show this coincidence, 
let us consider the general properties of  $V$ and $H_{\rm XY}$. 
The matrix $\sum_{j=1}^NK_{j}^{v*}\sigma_{j}^z$ is diagonal, 
and hence all the elements of 
$\exp[\sum_{j=1}^NK_{j}^{v*}\sigma_{j}^z]$ 
are non-negative. 
When $K_j^h\geq 0$, 
the elements of $\sum_{j=1}^NK_{j}^{h}\sigma_{j}^x\sigma_{j+1}^x$ 
are non-negative 
and hence the elements of $\exp[\sum_{j=1}^NK_{j}^{h}\sigma_{j}^x\sigma_{j+1}^x]$ 
are also non-negative. 
Therefore, in this case, all the matrix elements of $V$ are non-negative. 
Because the Ising interactions are written as  
$\sigma_{j}^x\sigma_{j+1}^x
=s_{j}^+s_{j+1}^++s_{j}^-s_{j+1}^-
+s_{j}^+s_{j+1}^-+s_{j}^-s_{j+1}^+$, 
the matrix $V$ becomes block-diagonal and contains two irreducible block elements, 
one is $V^{(1)}$ which operates on the bases states with even number of up spins 
and the other is $V^{(2)}$ which operates on the bases states with odd number of up spins: 
\begin{eqnarray}
V=
\left[
\begin{array}{cc}
V^{(1)} & O \\
O & V^{(2)}
\end{array}
\right]. 
\nonumber
\end{eqnarray}
The block element $V^{(1)}$ is irreducible 
because $(V^{(1)})^m$ is irreducible for sufficiently large $m$. 
(This comes from the fact that $(V^{(1)})^m$ cannot be irreducible if $V^{(1)}$ is reducible.) 
Similarly $V^{(2)}$ is also irreducible. 
Let $\Phi_{\rm I}^{(1)}$ and $\Phi_{\rm I}^{(2)}$ 
be the eigenstates of $V$ 
for the maximum eigenvalue of $V^{(1)}$ and $V^{(2)}$, respectively. 
These are expressed as 
\begin{eqnarray}
\Phi_{\rm I}^{(1)}=
\left[
\begin{array}{c}
\phi_{\rm I}^{(1)}  \\
{\bf 0}
\end{array}
\right],
\hspace{0.3cm}
\Phi_{\rm I}^{(2)}=
\left[
\begin{array}{c}
{\bf 0}\\
\phi_{\rm I}^{(2)} 
\end{array}
\right],
\nonumber
\end{eqnarray}
where $\phi_{\rm I}^{(1)}$ and $\phi_{\rm I}^{(1)}$ are vectors with $2^N/2$ elements, 
and ${\bf 0}$ is the zero vector with $2^N/2$ elements. 
From the Perron-Frobenius theorem, 
the elements of $\phi_{\rm I}^{(1)}$ and $\phi_{\rm I}^{(2)}$ 
are non-negative. 
Because the two block elements $V^{(1)}$ and $V^{(2)}$ are irreducible, 
$\phi_{\rm I}^{(1)}$ and $\phi_{\rm I}^{(2)}$ 
are non-degenerate eigenstates of $V^{(1)}$ and $V^{(2)}$, respectively, 
and the elements of $\phi_{\rm I}^{(1)}$ and $\phi_{\rm I}^{(2)}$ 
are all positive. 

Next let us consider the Hamiltonian $H_{\rm XY}$ with $J_j^k>0 \:(k=x, y)$. 
When we consider a matrix $-H_{\rm XY}+cI$,  
where $I$ is the unit matrix and $c$ is a positive and sufficiently large constant.  
The eigenstates of $-H_{\rm XY}+cI$ 
are also the eigenstates of $H_{\rm XY}$.  
In particular, the ground state of the Hamiltonian $H_{\rm XY}$ 
is the eigenstate for the maximum eigenvalue of $-H_{\rm XY}+cI$. 
Because the XY interactions are written as 
\begin{eqnarray}
&&J_j^x\sigma_{j}^x\sigma_{j+1}^x+J_j^y\sigma_{j}^y\sigma_{j+1}^y=
\nonumber\\
&&(J_j^x-J_j^y)(s_{j}^+s_{j+1}^++s_{j}^-s_{j+1}^-)
+(J_j^x+J_j^y)(s_{j}^+s_{j+1}^-+s_{j}^-s_{j+1}^+), 
\nonumber
\end{eqnarray}
the Hamiltonian $H_{\rm XY}$ is block-diagonalized into two blocks 
with even and odd number of up spins.    
From (\ref{cond-JJ}), the interaction constants satisfy $J_j^x>J_j^y$, 
and hence all the matrix elements of $-H_{\rm XY}+cI$ are non-negative 
and the two block elements are irreducible. 
Let $\Phi_{\rm XY}^{(1)}$ and $\Phi_{\rm XY}^{(2)}$ 
be the eigenstate of $H_{\rm XY}$ 
for the lowest energy eigenvalue of even and odd block element, respectively. 
These are expressed as 
\begin{eqnarray}
\Phi_{\rm XY}^{(1)}=
\left[
\begin{array}{c}
\phi_{\rm XY}^{(1)}  \\
{\bf 0} 
\end{array}
\right],
\hspace{0.3cm}
\Phi_{\rm XY}^{(2)}=
\left[
\begin{array}{c}
{\bf 0} \\
\phi_{\rm XY}^{(2)} 
\end{array}
\right]. 
\nonumber
\end{eqnarray}
It follows that 
the elements of $\phi_{\rm XY}^{(1)}$ and $\phi_{\rm XY}^{(2)}$ 
are all positive. 

The matrices $V$ and $H_{\rm XY}$ are both Hermitian,   
and hence their common eigenstates form 
(or can be arranged to form) an orthogonal basis set.  
Therefore, any two eigenstates of $V$ and $H_{\rm XY}$ 
are orthogonal or identical, or belong to a degenerate eigenspace. 
Because all the coefficients of 
$\phi_{\rm I}^{(1)}$ and $\phi_{\rm XY}^{(1)}$ are all positive, 
$\Phi_{\rm I}^{(1)}$ and $\Phi_{\rm XY}^{(1)}$ cannot be orthogonal.  
Considering the fact that 
$\phi_{\rm I}^{(1)}$ and $\phi_{\rm XY}^{(1)}$ 
are non-degenerate eigenstates of the corresponding block elements, respectively, 
it follows that 
$\Phi_{\rm I}^{(1)}=\Phi_{\rm XY}^{(1)}$ except overall constant.  
Similarly,  we obtain 
$\Phi_{\rm I}^{(2)}=\Phi_{\rm XY}^{(2)}$ 
except overall constant.  

In the cases with $K_j^h< 0$ and $J_j^k<0 \:(k=x, y)$ for some $j$, 
one can introduce a rotation of $\pi$ around the $z$ axis in the spin space, 
and the problem is mapped to one with $K_j^h>0$ and $J_j^k>0 \:(k=x, y)$, 
which yields the same result.

The remaining problem is 
to identify the eigenstate for the maximum eigenvalue of $H_{\rm I}$ 
and the ground state of $H_{\rm XY}$. 
For the periodic case 1-1) with 2-1), 
it has been shown in\cite{71Suzuki} through a direct diagonalization by fermion operators 
that the eigenstate for the maximum eigenvalue of $V$ is $\Phi_{\rm I}^{(1)}$, 
and that the ground state of $H_{\rm XY}$ is $\Phi_{\rm XY}^{(1)}$.

In the case of 1-1) with 2-2), the XY model has free ends. 
The Hamiltonian $H_{\rm XY}$ is separated into two sectors, 
one with even number of fermions 
( i.e. the block-element with even number of up spins), 
and the other with odd number of fermions 
( i.e. the block-element with odd number of up spins). 
It is most clearly seen in the calculation by Niemeijer\cite{67Niemeijer} that 
the energy eigenvalues and the eigenvectors are obtained 
from an eigenvalue problem for a symmetric matrix. 
Hence the XY model with free boundary conditions can be solved, 
and the ground state is unique and found in the even sector. 

Schultz et al.\cite{64SchultzMattisLieb} 
diagonalized the two-dimensional Ising model 
with free boundary conditions in one direction. 
The maximum eigenvalue of the transfer matrix  
was found in the sector with even number of fermions 
( i.e.  in the block-element with even number of up spins), 
and was non-degenerate above $T_c$. 
Below $T_c$, 
the maximum eigenvalue of the odd sector 
becomes degenerate with that of the even sector in the thermodynamic limit. 

Consequently, 
we again find, for the case 1-1) with 2-2), 
that the eigenstate for the maximum eigenvalue of $V$ is $\Phi_{\rm I}^{(1)}$, 
and the ground state of $H_{\rm XY}$ is $\Phi_{\rm XY}^{(1)}$, 
where  $\Phi_{\rm I}^{(1)}=\Phi_{\rm XY}^{(1)}$ 
except overall constant.

The cases with $K_j^h< 0$ and $J_j^k<0 \:(k=x, y)$ for some $j$ 
can be mapped to the cases with $K_j^h>0$ and $J_j^k>0 \:(k=x, y)$  
by a rotation  of $\pi$ around the $z$ axis in the spin space.   
Because the energy eigenvalues are invariant by the rotation, 
the results for 1-1) with 2-1) or 2-2) 
are still valid in the cases 1-2) and 1-4) with 2-1) or 2-2). 

Finally, let us consider the cases 1-3) and 1-5)  with 2-1) or 2-2).  
The system with $K_j^h>0$ can be mapped to the system with $K_j^h<0$,  
and the rectangular Ising model with $K_j^v<0$ and $K_j^h<0$ for all $j$ 
can be mapped to that with $K_j^v>0$ and $K_j^h>0$.  
It is easy to check that 
the corresponding XY models are also mapped each other by a gauge transformation. 
Therefore, we can assume that $K_j^v>0$ and $K_j^h>0$ for all $j$. 
Then the elements of $V_1$ and $V_2$ are all non-negative 
because $K_j^{v*}$ are real and $K_j^h$ are positive. 
Without loss of generality, 
one can assume that $\tau_j>0$ in (\ref{cond-JJ}) and then
the interactions $J_j^x$ and $J_j^y$ satisfy $J_j^x-J_j^y>0$ and $J_j^x+J_j^y>0$. 
All the non-zero elements of $V$ and $-H_{\rm XY}+cI$ remain non-zero 
as long as $K_j^v$ and $K_j^h$ remain real, finite and non-zero. 

From the Perron-Frobenius theorem, 
the maximum eigenvalue of each block element is non-degenerate. 
When we assume that the eigenvalues are continuous in terms of the coupling constants,   
this leads to the result that 
level crossings do not occur  in each block element 
as long as all the non-zero matrix elements remain non-zero.  
Therefore $\phi_{\rm I}^{(1)}=\phi_{\rm XY}^{(1)}$ is the eigenstate 
for the maximum eigenvalue of the even block element, 
and $\phi_{\rm I}^{(2)}=\phi_{\rm XY}^{(2)}$ is 
that of the odd block element, 
for real, finite and non-zero $K_j^v$ and $K_j^h$. 
The state $\Phi_{\rm I}^{(1)}=\Phi_{\rm XY}^{(1)}$ 
is the eigenstate for the maximum eigenvalue of $V$ 
as well as the ground state of $H_{\rm XY}$,  
at least in a neighborhood of the uniform point 1-1). 

The correlation functions of the two-dimensional Ising model 
and those in the ground state $\Phi_{\rm XY}^{(1)}$ of the one-dimensional XY model 
are related to each other. 
Let $f(\sigma)$ be a product of the Pauli operators $\sigma_{j}^x$. 
When one take the limit $M\to\infty$ in the equation
\begin{eqnarray}
\frac{{\rm Tr}f(\sigma)(V_1V_2)^M}{{\rm Tr}(V_1V_2)^M}
=
\frac{{\rm Tr}f(\sigma)V_1^{1/2}V^MV_1^{-1/2}}{{\rm Tr}V^M},
\nonumber
\end{eqnarray}
one obtains the relation of the expectation values as 
\begin{eqnarray}
\langle f(\sigma)\rangle_{\rm I}
=
\langle V_1^{-1/2}f(\sigma)V_1^{1/2}\rangle_{\rm XY}, 
\nonumber
\end{eqnarray}
where $\langle f(\sigma)\rangle_{\rm I}$ is the expectation value of a spin product 
on the same row of the Ising model. 
In particular, the two-spin correlation functions satisfy the relation 
\begin{eqnarray}
\langle \sigma_{ij}^x\sigma_{ik}^x\rangle_{\rm I}
=
\cosh K_j^v\cosh K_k^v
\langle \sigma_{j}^x\sigma_{k}^x\rangle_{\rm XY}
-\sinh K_j^v\sinh K_k^v
\langle \sigma_{j}^y\sigma_{k}^y\rangle_{\rm XY},
\nonumber
\end{eqnarray}
where we used the fact that 
all the elements of $\Phi_{\rm XY}^{(1)}$ are real, 
and hence $\langle \sigma_{j}^x\sigma_{k}^y\rangle_{\rm XY}=0$ for $j\neq k$. 

As shown in \cite{71Suzuki}, 
the critical point $T_c$ of the two-dimensional Ising model 
corresponds to the critical field $H_c$ of the one-dimensional XY model,      
and two quivalent models show the same critical singularities.

However, especially in the cases with random coupling parameters, 
complex critical behavior can occur depending on the distribution function for $\{K_j^v\}$. 
It is known that the Griffiths-McCoy phase, 
which was originally found in two-dimensional random Ising models,\cite{69McCoy}\cite{69Griffiths}
appears
in the one-dimensional random transverse Ising model,\cite{95Fisher} 
and one-dimensional XY model 
with random interactions and random fields\cite{96McKenzie}.  
As mentioned above, 
it is also found\cite{87ShankarMurthyRC}\cite{87ShankarMurthy} that 
the Griffiths-McCoy phase appears 
in the two-dimensional Ising model with random horizontal interactions, 
which corresponds to our case 1-5) with 1-1). 
The relations derived in this letter 
shows the exact equivalence of certain kind of two-dimensional random Ising models 
and the one-dimensional random XY models.

\end{document}